\documentclass[12pt]{article}

\usepackage{amsmath,amssymb}

\usepackage[all]{xy}
\usepackage{color}
\usepackage{fullpage}
\usepackage[vcentermath]{youngtab}
\usepackage{amsfonts}
\usepackage{amsbsy}
\usepackage{amsthm}
\usepackage{graphicx}
\usepackage{epstopdf}

\def\Z{\mathbb{Z}}

\def\O{\mathcal{O}}

\def\tr{\mathop{\mathrm{tr}}}
\def\tra{\mathop{\mathrm{Tr}}}
\def\trad{\mathop{\mathrm{tr}_a}}
\def\rr{\tr R^2}
\def\talpha{\tilde{\alpha}}

\begin{document}

\setcounter{page}{1}

\baselineskip16pt

\begin{titlepage}

\begin{flushright}
\hfill MIT-CTP-4046\\
\end{flushright}
\vspace{13 mm}
\vspace*{0.6in}

\begin{center}

{\Large \bf String Universality in Six Dimensions}

\end{center}

\vspace{7 mm}

\begin{center}

Vijay Kumar and Washington Taylor

\vspace{3mm}
{\small \sl Center for Theoretical Physics} \\
{\small \sl Massachusetts Institute of Technology} \\
{\small \sl Cambridge, MA 02139, U.S.A.} \\
{\small {{\tt vijayk, wati} {\rm at} {\tt mit.edu}}}\\
\end{center}

\vspace{8 mm}

\begin{abstract}
In six dimensions, cancellation of gauge, gravitational, and mixed 
anomalies strongly constrains the set of quantum field theories
  which can be coupled consistently to gravity.  We
  show that for some classes of six-dimensional supersymmetric gauge
  theories coupled to gravity, the 
  anomaly cancellation conditions are equivalent to
  tadpole cancellation and other constraints on the matter content of
  heterotic/type I compactifications on K3.  In these cases, all
  consistent 6D supergravity theories have a realization in string
  theory.  We find one 
  example which may arise from a novel string compactification, and we
  identify a new infinite family of models satisfying anomaly
  factorization.  We find, however, that this infinite family of
  models, as well as other infinite families of models previously
  identified by Schwarz are pathological. We suggest that it may be feasible to
  demonstrate that there is a string theoretic realization of all
  consistent six-dimensional supergravity theories which have
  Lagrangian descriptions with arbitrary gauge and matter content.  We
  attempt to frame this hypothesis of string universality as a
  concrete conjecture.
\end{abstract}

\vspace{1cm}
\begin{flushleft}
June 2009
\end{flushleft}
\end{titlepage}
\newpage

\vspace{-6in}

\newpage

\tableofcontents

\section{Introduction}

The vast plethora of apparently consistent string vacua has posed a
challenge for string theory since the early days of the subject.
Despite several decades of work, there is still no hint
of any dynamical mechanism which might select one particular string
vacuum solution.  It seems clear at this point that there is an enormous
range of distinct string compactifications giving rise to field
theories in four dimensions through flux compactifications and related
constructions \cite{flux-review}.  In fact, it has been shown that
there are discrete infinities of such string solutions \cite{IIA}.
Nonetheless, we are far from a full understanding of the range of
possibilities, and there is an even larger set of low-energy field
theories in four dimensions for which no obstruction is known to a
consistent coupling to quantum gravity.  The space of apparently
consistent low-energy theories which do not admit a realization in
string theory was dubbed the ``swampland'' by Vafa \cite{swampland}.  In
four dimensions this swampland seems quite vast given the current
state of our knowledge.

In six dimensions, the situation is quite different from that in four
dimensions.  The constraints posed by cancellation of gauge,
gravitational, and mixed anomalies in dimensions $4k + 2$ are very
stringent for theories with chiral (1, 0) supersymmetry
\cite{Alvarez-Gaume-Witten, Schwarz}.  While the range of allowed
gauge groups and massless field content in six dimensions is
apparently quite large (see \cite{Avramis-Kehagias} for an extensive
list and further references), the range of available string
constructions through heterotic \cite{gsw}, F-theory
\cite{Morrison-Vafa, geometric-singularities, Aspinwall-Morrison, Bershadsky-Vafa},
and other constructions is comparably extensive.  The anomaly
constraints essentially arise from an index theorem in the noncompact
space-time dimensions, while the constraints on string constructions
come from a combination of a tadpole constraint (related to an index
in the compact space) and another application of the index theorem,
describing the number of massless fields in a certain brane
configuration or geometry.  The anomaly constraints in six dimensions
are closely related to the constraints on string compactifications \cite{Blum1,
Blum2, Leigh, Park, aldazabal, Bianchi, Uranga, Grassi-Morrison}.

In this paper we consider some specific classes of six-dimensional
field theories.  We find that in these classes of models, the anomaly
cancellation conditions in six dimensions are essentially equivalent
to the tadpole and matter field constraints in string
compactifications.  In the previous literature \cite{Schwarz} infinite
families of models have been discovered which satisfy anomaly
factorization, a necessary condition for anomaly cancellation.  
No string compactification is known which corresponds to the theories in this
infinite family.
We find a new infinite class of theories, and show that these models, as well
as those found in \cite{Schwarz}, are pathological.
Motivated by the close
relationship between the anomaly cancellation conditions in six
dimensions and the constraints on string compactifications, and the
relative paucity of examples of six-dimensional supergravity theories
which are apparently consistent and yet lack a string description, we
suggest that it may be within reach at the current time to give a
definitive proof that string theory realizes all consistent, supersymmetric 
theories of quantum gravity in six dimensions.  Furthermore, we expect that
this class of consistent models can be determined entirely from  
the low-energy Lagrangian. We attempt to provide a
framework for discussing and moving towards a proof of this hypothesis
of string universality through a concrete conjecture.

In Section \ref{sec:examples}, we give examples of classes of models
satisfying the anomaly cancellation conditions, and show that almost
all of the models in these classes which appear to be consistent can
be realized through existing string constructions.  We discuss an
apparent problem arising from the gauge kinetic terms of a large
family of models which do not have string realizations. We also show that for
the specific choice of gauge group $SU(N)$ with arbitrary matter content, there
are only finitely many theories that can be consistently coupled to 
supergravity. In section
\ref{sec:conjecture} we state the conjecture of string universality in
six dimensions more precisely, and discuss various aspects of this
hypothesis.  Section \ref{sec:conclusions} contains concluding
remarks.

\section{Examples of 6D models}
\label{sec:examples}

A wide range of string constructions of supersymmetric 6D models have
been studied in the literature.  Some of the first models of this type
were compactifications of type I or heterotic strings on K3 using line
bundles \cite{gsw}.  More general vector bundle compactifications were
described using F-theory in \cite{Kachru-Vafa, Vafa-f, Morrison-Vafa}.  
Compactifications on orbifolds give rise to a further range of
theories \cite{6d-orbifolds, o2}, and
gauge group enhancement from small instantons on singularities
gives rise to a rich variety of six-dimensional theories which
have been described in F-theory language \cite{Aspinwall-Morrison}.

Some work has also been done in characterizing 6D field theories which
exhibit cancellation of gauge, gravitational, and mixed anomalies,
independent of string theoretic considerations \cite{6D-anomalies, 6D-theories,  Sagnotti-others, Sagnotti}.  In \cite{Schwarz}, 
several models (including two infinite families) were constructed which satisfy
anomaly factorization.  In \cite{Avramis-Kehagias}, a systematic
search for models satisfying anomaly cancellation was carried out, in
the case where the gauge group has one or two factors, of which at
most one factor is one of the classical groups $SU(N), SO(N)$, or
$Sp(N)$.  Of the solutions of the anomaly cancellation conditions
found so far, many correspond to known string constructions.  Others
do not.

In this section we consider two classes of models.  First, in
Subsection~\ref{sec:uo}, we look at models whose gauge group has the
form $SU(N) \times SO(M)$, with certain types of allowed matter
fields.  We find that the anomaly cancellation conditions in this case
are essentially equivalent to the conditions giving the tadpole
constraint and matter content of a class of string compactifications,
so that all of the models in this class which can be consistently
coupled to gravity have a realization in string theory (with one
possible exception). The same qualitative results hold with multiple
factors of $SU(N)$ and a single $SO(M)$.  We then consider in
Subsection~\ref{sec:infinite} several infinite families of models
satisfying anomaly factorization, one of which arises in our analysis
of the gauge group $SU(N) \times SO(M)$, the others of which were
found in \cite{Schwarz}.  We find that these theories have wrong-sign
kinetic terms at a general point in moduli space and seem to be
pathological.  Finally, we conclude this section with a demonstration
in Subsection~\ref{sec:existence-infinite} that there are only finitely many 
anomaly-free models with gauge group $SU(N)$, with no
restrictions whatsoever on allowed matter representations.


\subsection{Gauge group $SU(N) \times SO(M)$}
\label{sec:uo}

As an example of the correspondence between anomaly-free low-energy
models and string constructions, we consider models with gauge group
$SU(N) \times SO(M)$.  To simplify the analysis we restrict to $N
\geq 4$ and $M \geq 4$.  A similar analysis can be done for smaller
values of $N, M$ with similar results, though the details are slightly
more complicated.  

The motivation for considering this particular class of models is that
we know of a simple string construction giving one set of models of
this type.  These models were constructed through heterotic
string compactifications using line bundles in \cite{gsw}, and studied
further in \cite{Honecker}.
In \cite{Kumar-Taylor}, we found a clean mathematical
characterization of the general class of line bundle models constructed in
\cite{gsw} in terms of lattice embedding theorems of Nikulin.  These
results demonstrated that for a certain parameterization of possible
models, all models could be constructed, in most cases in a unique
fashion, through K3 string compactifications.  Showing that all models
in some appropriate class which satisfy anomaly cancellation can be
realized through the parameterization of \cite{Kumar-Taylor} would
demonstrate string universality for this subclass of models, and
furthermore prove uniqueness for the string realization of these
models, up to some discrete redundancies.  

The models constructed in \cite{gsw, Kumar-Taylor} are given by
turning on U(1) fluxes in the type I compactification on K3.  The
resulting low-energy supersymmetric six-dimensional theories have rank
16 gauge groups of the form
\begin{equation}
U(N_1)\times U(N_2) \times  \cdots
\times U(N_k)\times
SO(32-2 \sum_{i} N_i) \,,
\label{eq:big-group}
\end{equation}
with matter fields in the bifundamental representation connecting
various components of the gauge group and in the antisymmetric
representations of the components $U(N_i)$.  The abelian $U(1)$
factors in the groups $U(N_i)$ are anomalous and acquire masses by the
Green-Schwarz mechanism \cite{afiru, imr, berkooz}. As a result, the
gauge group contains $SU(N_i)$ factors with fewer neutral hypermultiplets.

We consider theories with the gauge group $SU(N)
\times SO(M)$ with $B$ massless hypermultiplets in the
bifundamental representation $(N, M)+(\bar{N},M)$ and $A$
massless hypermultiplets in the two-index antisymmetric representation of
$SU(N)$, ${\scriptsize \yng(1,1)}+\mbox{c.c}$ \footnote{In six dimensions, hypermultiplets 
consist of two complex scalars, and can only be charged in a real representation
of the gauge group. Henceforth, the complex conjugate is implicit if the
representation is complex and we will drop it in our notation.}. In addition,
these theories include the supergravity multiplet, one tensor multiplet and an
arbitrary number of neutral hypermultiplets.
The models we consider here are generalizations of the $k=1$ models in \eqref{eq:big-group}.  
We could also include hypermultiplets in the fundamental
representations of these gauge groups; this would give a larger class
of solutions including those realized by Higgsing the models under
consideration here.  For simplicity of the analysis, however, we
restrict attention to the bifundamental and antisymmetric
representations found in the models of \cite{gsw, Kumar-Taylor}.  This
analysis naturally complements that of \cite{Avramis-Kehagias}, where
products of two groups were considered with zero or one classical
factor.

In the heterotic/type I line bundle construction of \cite{gsw,
Kumar-Taylor}, the parameters $N, M, A, B$ satisfy the following
constraints
\begin{eqnarray}
2N + M & = &  32  \label{eq:9-tadpole}\\
N \tau & = &   24\label{eq:5-tadpole}\\
A & = &  -2 + 4 \tau\label{eq:a-number}\\
B & = &  -2 + \tau\label{eq:b-number}
\end{eqnarray}
where $\tau$ is an integer parameter (in the notation of
\cite{Kumar-Taylor}, $\tau = -f^2$ encodes the instanton number on a
stack of $N$ D9-branes). 
Conditions (\ref{eq:9-tadpole}) and (\ref{eq:5-tadpole}) arise from
the D9-brane and D5-brane tadpole conditions, while
(\ref{eq:a-number}) and (\ref{eq:b-number}) are computed from the
index theorem on the D9-brane world volume giving the number of
massless fields in each representation of the gauge group. 
These theories have 20 neutral 
hypermultiplets coming from the K3 moduli. One of these
hypermultiplets is eaten by the (anomalous) $U(1)$ vector multiplet through the
Green-Schwarz mechanism, leaving 19 neutral hypermultiplets. 
Our goal now is to show that (\ref{eq:9-tadpole}-\ref{eq:b-number}) are
equivalent to the anomaly cancellation conditions for the associated
theories in  six dimensions.

The gravitational and gauge anomalies in six-dimensional
supersymmetric theories arise from chiral fermions in hyper and vector
multiplets of the supersymmetry algebra (as well as from the gravitino
and self-dual and anti-self-dual 2-form fields).  Anomaly cancellation
in six-dimensional models is described in \cite{gsw}, and analyzed
further in \cite{Erler}, which gives a concise summary of the anomaly
cancellation conditions.  We restrict attention here to theories
admitting a Lagrangian description, so that the number of tensor
multiplets is $n_T = 1$.  For such models, cancellation of the $R^4$
anomaly requires the relation
\begin{equation}
n_H-n_V = 244,
\label{eq:grav-anomaly}
\end{equation}
where $n_H$ and $n_V$ are the number of hyper- and vector multiplets
respectively.  Since some number of hypermultiplets can be in the
trivial representation of the gauge group, this is not {\it a priori}
a strong constraint on allowed models; we will use this to compute the
number of hypermultiplets transforming as singlets in anomaly-allowed
low-energy theories and compare with the expected number of 19 (20-1) neutral
hypers which come from the closed string sector in general string
compactifications.

The cancellation of the remaining gravitational, gauge, and mixed
anomalies requires that the anomaly polynomial $I$ expressing these
anomalies can be written in factorized form as required by the
Green-Schwarz mechanism.  The anomaly polynomial for the models we
consider here is proportional to
\cite{Erler, Schwarz}
\begin{eqnarray}
I & = &(\rr)^2 + \frac{1}{6}\rr(\tra \;F_2^2 - BN \tr
F_2^2) + \frac{1}{6}\rr(\tra \; F_1^2  \nonumber \\
& & - A \trad \;F_1^2-BM\tr F_1^2)  - \frac{2}{3} (\tra \;F_2^4 - BN \tr F_2^4) \nonumber \\
&  & - \frac{2}{3} (\tra \; F_1^4 - A\trad \;F_1^4-BM \tr F_1^4) + 4B \tr F_1^2 \tr F_2^2 
\nonumber
\end{eqnarray}
where Tr denotes the trace in the adjoint representation, tr denotes
the trace in the fundamental representation, and $\trad$ denotes the
trace in the antisymmetric representation of $SU(N)$.

Using the trace identities  \cite{Erler}
\begin{equation}
\begin{array}{crcl}
SU(N): & 	\tra \; F^2 & = & 2N \tr F^2 \\
 & 		\trad \; F^2 & = & (N-2) \tr F^2 \\
&			\tra \; F^4 & = & 2N \tr F^4 + 6 (\tr F^2)^2 \\
&			\trad \; F^4 & = & (N-8) \tr F^4 + 3 (\tr F^2)^2 \\
& & &     \\
SO(N):  &	\tra \; F^2 & = & (N-2)\tr F^2 \\
 & 			\tra \; F^4 & = & (N-8) \tr F^4 + 3 (\tr
F^2)^2 
\end{array}
\end{equation}
the polynomial simplifies to
\begin{eqnarray}
I &= & X^2 + \frac{1}{6}(M-2- BN)XZ + \frac{1}{6}(2N-A(N-2)-BM) XY 
\nonumber\\
& & \hspace*{0.1in}-2Z^2 +  (2A-4)Y^2 + 4BYZ \nonumber \\ 
&  &\hspace*{0.1in}- \frac{2}{3}(M-8-BN)\tr F_2^4 - \frac{2}{3} (2N-A(N-8)-BM)\tr F_1^4 
\end{eqnarray}
where $X:=\rr$, $Y:= \tr F_1^2$, $Z := \tr F_2^2$.  We assume that $N
\geq 4$ and $M \geq 4$. Both $\tr F^4$ terms are then present for all relevant
values of $N,M,A,B$. For anomaly cancellation, the polynomial must
factorize, which requires any $\tr F^4$ terms to be absent. This gives
two equations
\begin{eqnarray}
M-BN & = & 8  \nonumber\\
2N-AN+8A-BM & = & 0 \label{eq:f4-conditions}
\end{eqnarray}
Substituting these conditions in $I$, we have
\begin{eqnarray}
I  =  X^2+(2A-4)Y^2-2Z^2 - AXY+4BYZ+XZ
\label{eq:simplified}
\end{eqnarray}
For anomaly cancellation, $I$ must have the factorized form
\begin{eqnarray}
I =  (X -\alpha_1 Y-\alpha_2Z) (X -\tilde{\alpha}_1
Y-\tilde{\alpha}_2Z) \label{eq:factorized-general}
\end{eqnarray}
The polynomial (\ref{eq:simplified}) can be factorized as above only if the
following system of equations has a solution.
\begin{eqnarray}
\alpha_1+\talpha_1=A, \hspace*{0.1in} & &
\hspace*{0.1in}  \alpha_1\talpha_1 = 2A-4  \label{eq:a1}\\
\alpha_2+\talpha_2=-1,\hspace*{0.1in} & &
\hspace*{0.1in}\alpha_2\talpha_2=-2 
\label{eq:a2}\\
\alpha_1\talpha_2 + \alpha_2\talpha_1   &=   &4B  \label{eq:a12}
\end{eqnarray}

There are two solutions to the pair of equations (\ref{eq:a2}), namely
$\alpha_2 = 1, \tilde{\alpha}_2 = -2$ and
$\alpha_2 = -2, \tilde{\alpha}_2 = 1$.
Since the factorized form is symmetric, without loss of generality we
take
\begin{equation}
\alpha_2 = 1, \quad \tilde{\alpha}_2 = -2 \,.
\label{eq:a2v}
\end{equation}
There are also two solutions to the pair of equations
(\ref{eq:a1}). The possibilities are
\begin{eqnarray}
{\rm {\bf Case\ 1}} & \hspace*{0.1in} &  \alpha_1 = 2, \quad
 \tilde{\alpha}_1 = A-2 \label{eq:cases}\\
{\rm {\bf Case\ 2}} & \hspace*{0.1in} &  \alpha_1 = A-2, \quad
 \tilde{\alpha}_1 = 2\nonumber
\end{eqnarray}
In all known six-dimensional string constructions giving gauge groups
with a $U(N)$ factor, the associated value of the parameter $\alpha$
in the anomaly factorization condition is $\alpha = 2$
\cite{Erler, iu}.  We now discuss these two cases in turn.
\vspace*{0.05in}

\noindent
{\bf Case 1}: $\alpha_1 = 2, \quad
 \tilde{\alpha}_1 = A-2$

In this case, (\ref{eq:a12}) combined with (\ref{eq:a2v})
gives
\begin{equation}
\alpha_1\talpha_2 + \alpha_2\talpha_1
= A -6 = 4B   \,.
\label{eq:ab1}
\end{equation}
This equation is precisely the
relation between the numbers of matter fields dictated by
(\ref{eq:a-number}), (\ref{eq:b-number})
\begin{eqnarray}
A & = & -2+4\tau  \nonumber\\
B & = & -2 + \tau\label{eq:ab-numbers}
\end{eqnarray}
where we can replace
$B$  with the integer parameter $\tau$ through $B = \tau -2$.  Thus,
given cancellation of $F^4$ terms,
factorization of the anomaly 
is equivalent to the constraints on the matter fields 
in the ten-dimensional compactification.  Returning to the $F^4$
anomaly cancellation conditions, replacing $A$ and $B$ with $\tau$
through (\ref{eq:ab-numbers})
gives
\begin{eqnarray}
M + 2 N -\tau N -8 & = &  0 \label{eq:nmt-1}\\
2 N + (2-4 \tau) N -16+ 32 \tau + (2-\tau) M & = &  0 \,.   \label{eq:nmt-2}
\end{eqnarray}
Subtracting  (\ref{eq:nmt-2}) from
twice  (\ref{eq:nmt-1}) gives
\begin{equation}
\tau (M+ 2 N -32) =0\,.
\label{eq:9-again}
\end{equation}
Since $B  \geq 0$, $\tau \geq 2$, (\ref{eq:9-again}) is equivalent
to the D9-brane tadpole cancellation condition (\ref{eq:9-tadpole}).
Imposing this condition on (\ref{eq:nmt-1}) gives
\begin{equation}
\tau N = 24
\end{equation}
which is precisely the D5-brane tadpole cancellation condition
(\ref{eq:5-tadpole}).
Thus, in Case 1, the anomaly cancellation condition is precisely
equivalent to the tadpole cancellation and matter contact constraints
in a class of string vacuum constructions.
\vspace*{0.05in}

\noindent
{\bf Case 2}:  $\alpha_1 = A-2, \quad \tilde{\alpha}_1 = 2$

In this case, analogous to (\ref{eq:ab1}) we have the relation
$A+2B=3$.  The only solutions to this for nonnegative
$A, B$ are $(A,B)=(3,0),(1,1)$. 
For $(A, B)=(3,0)$
substituting in
(\ref{eq:f4-conditions}) gives the solution
\begin{equation}
(N,M,A,B)=(24,8,3,0)\,.
\end{equation}
For $(A, B) = (1, 1)$, there is an infinite family of solutions
\begin{equation}
(N,M,A,B)= (N,N+8,1,1)\,.
\end{equation}

\begin{table}
\centering
\begin{tabular}{|c|c|c|c|c|}
\hline
& Gauge Group & Anti $U(N)$ & $(N,M)$ & Neutral Hypers \\
\hline
1 &$SU(4)\times SO(24)$ &22& 4& 19 \\
2 &$SU(6)\times SO(20)$ &14& 2& 19 \\
3 &$SU(8)\times SO(16)$ &10& 1& 19 \\
4 &$SU(12)\times SO(8)$ & 6& 0& 19 \\
\hline
5 &$SU(24)\times SO(8)$ & 3& 0& 19 \\
6 &$SU(N) \times SO(N+8)$& 1&1& 271 \\
\hline
\end{tabular} \label{t:solutions}
\caption{Models solving anomaly cancellation conditions 
for
 $N\geq 4$, $M \geq 4$} 
\end{table}

The anomaly free theories are shown in Table~\ref{t:solutions}. 
Some comments are relevant regarding this list of models.  
In order to satisfy (\ref{eq:grav-anomaly}), which is the condition
for the absence of purely gravitational anomalies, we add gauge
singlet hypermultiplets.  The number of neutral hypermultiplets in
each case is computed using the gravitational anomaly cancellation
condition (\ref{eq:grav-anomaly}), and included in
Table~\ref{t:solutions}.  In all cases except the infinite family (6),
the number of hypermultiplets is 19, as expected from a
string compactification on K3.

We have found models (1-4) which have the same forms as known
compactifications on K3.  Extending the analysis for smaller $N$ gives
additional models at $N = 3, 2$, associated with the compactifications
found in \cite{gsw,Kumar-Taylor}.  The case $N = 8$ is not found
through the line bundle construction of \cite{gsw, Kumar-Taylor}.  In
this case, in the type I string picture, a flux is turned on in the
diagonal $U(1)$ on a stack of 8 D9-branes with instanton number 24.
This gives odd $\tau = 3$, and cannot be decomposed into a sum of
distinct line bundle contributions on the individual D9-branes, but is
a perfectly satisfactory string compactification.  In the case $N =
12$, some care may also be needed as in this case the instantons wrap a
cycle with $f^2 = -2$ which has vanishing size.  Gauge group
enhancement may occur in this situation, analogous to the situation
described in \cite{Aspinwall-Morrison}, so a more careful analysis of
this case may be warranted.

We are not aware of an explicit string construction that realizes
Theory (5), but it seems reasonable to assume that such a realization
exists. The theory has 19 neutral hypermultiplets, typical of K3
compactifications.  The unusual value of $\alpha_1 = 1$, which differs
from the value $\alpha = 2$ typical of string constructions with a
$SU(N)$ gauge group component \cite{Erler, iu}, may give some hint for
a string construction of this model.  It is tempting to associate the
$SU(24)$ with 24 D5-branes which saturate the tadpole on K3, but that theory has
gauge group $Sp(24) \times SO(32)$ \cite{small-instantons}, since D5-branes in
the type I theory carry a Symplectic group.


Thus, all but one of theories (1-5) have a string construction, and it
seems plausible that a string realization will also be found for the
last one (5).  The class of theories (6), on the other hand, seems
problematic. This corresponds to an infinite family of 6d gauge theories
with increasing rank. In a K3 compactification, there seems to be an
upper bound on the rank of the gauge group \cite{Aspinwall-Morrison},
and therefore it is questionable whether such an infinite family could ever
be realized as a string compactification. However, as we show in the
next section, this theory and other infinite families that have been
found by Schwarz have a common pathology which likely renders these 
theories inconsistent.

We have thus found that for this class of models there is a close
correspondence between the anomaly cancellation conditions in six
dimensions and the conditions determining the allowed gauge group and
matter content in a class of string compactifications.  A similar
analysis can be done for other classes of models.  We expect that in
many other situations the results will be similar.  For example, one
can go to ``$k$-stack'' models with gauge group $SU(N_1) \times \cdots
\times SU(N_k) \times SO(M)$.  In this case the equations have a
similar structure and give rise to similar constraints.  In fact, the
equations governing these models are virtually identical to the
analysis above, the only differences being that for each component
$SU(N_i)$ there are separate numbers of multiplets $A_i, B_i$ and
separate coefficients $\alpha_i, \tilde{\alpha}_i$ in the
factorization equation (\ref{eq:factorized-general}) for each $i$, as
well as a number of hypermultiplets $\beta_{ij}, \hat{\beta}_{ij}$ in the $(N_i,
\bar{N}_j), (N_i, N_j)$ representations.  Choosing the solution associated with
Case 1 in (\ref{eq:cases}) for all $i$ reproduces the equations for
line bundle models with $k$ stacks, though the connection with string
compactifications becomes more complicated at $k = 4$.  Choosing other
cases gives a small number of other models analogous to the $SU(24)
\times SO(8)$ model discussed above.  We leave further analysis of
these $k$-stack generalizations of the $SU(N) \times SO(M)$ models for
further work.

\subsection{Infinite families with factorized anomalies}
\label{sec:infinite}

The class of models (6) in Table~\ref{t:solutions} gives an apparent
infinite family of solutions to the anomaly cancellation equations.
In searching for six-dimensional models satisfying anomaly cancellation, Schwarz
discovered two other infinite families of models satisfying anomaly factorization 
\cite{Schwarz}. These infinite families are shown in
Table~\ref{t:infinite-families} along with (6) from 
Table~\ref{t:solutions}.

\begin{table}
\centering
\begin{tabular}{|c|c|c|}
\hline
Gauge Group & Matter content & Anomaly polynomial\\
\hline
$SU(N)\times SU(N)$ & $2({\tiny \yng(1)}, \bar{\tiny \yng(1)})$ & $(X-2Y+2Z)(X+2Y-2Z)$\\
\hline
$SO(2N+8)\times Sp(N)$ & $({\tiny \yng(1)},{\tiny \yng(1)})$ & $(X-Y+Z)(X+2Y-2Z)$\\
\hline
$SU(N) \times SO(N+8)$& $({\tiny \yng(1)},{\tiny \yng(1)})+ {\tiny \yng(1,1)}$  & $(X+Y-Z)(X-2Y+2Z)$\\
\hline
\end{tabular} \label{t:infinite-families}
\caption{Infinite families of anomaly-free 6d theories, where the anomaly
polynomial factorizes as shown. In each case, the number of neutral hypermultiplets can be calculated from the gravitational anomaly condition $n_H-n_V=244$. $X,Y,Z$ denote $\tr R^2, \tr F_1^2, \tr F_2^2$
respectively, where $F_1$ is the field strength of the first gauge group factor
and $F_2$ that of the second.}
\end{table}

It was shown by Sagnotti \cite{Sagnotti} (see also \cite{dmw}) that the coefficients
$\alpha_i,\talpha_i$ in the anomaly polynomial are related by supersymmetry to
the coefficients of the gauge field kinetic terms in the Lagrangian. For the
anomaly polynomial
\begin{eqnarray}
I & = & X_4 \tilde{X}_4 \nonumber \\
X_4 & = & \tr R^2 - \sum_i \alpha_i \tr F_i^2 \\
\tilde{X}_4 & = & \tr R^2 - \sum_i\talpha_i \tr F_i^2 \nonumber,
\end{eqnarray}
the gauge field kinetic terms in Einstein frame are
\begin{equation}
\mathcal{L}_{\mbox{{\scriptsize gauge}}} \propto - \sum_i \left(\alpha_i
e^{-\phi/2} + \talpha_i e^{\phi/2} \right) \tr F_i^{\mu\nu}F_{\mu\nu}^i
\end{equation}
Here $\phi$ is the scalar in the tensor multiplet. 

In all three infinite families shown in
Table~\ref{t:infinite-families}, $\alpha_1=-\alpha_2,
\talpha_1=-\talpha_2$. As a result, the gauge kinetic term of one of
the gauge group factors generically has the wrong sign, except at the
special value $\exp (\langle \phi \rangle)= - \alpha_1/\talpha_1 =
-\alpha_2/\talpha_2$.  This corresponds to the case when both kinetic
terms vanish and the 6d gauge coupling is infinite.  
The wrong-sign kinetic terms for at least one gauge group factor seem
to indicate an instability which renders these models inconsistent.
Thus, we assume that these models cannot be sensibly interpreted as
quantum supergravity theories in 6D with Lagrangian descriptions.  
Even more subtle ``wrong sign'' terms in low-energy Lagrangians have
been shown to lead to inconsistencies in the UV \cite{Allan-Nima}.
It
is possible, however, that there may be some unconventional way of
making sense of these theories.  
By analogy with ``ghost condensation''
\cite{nima}, it is possible that there is a phase transition to a new, 
Lorentz-violating vacuum in 
these infinite families; we are, however, interested in theories which are 
Lorentz
invariant. 
There is a point in the tensor multiplet moduli space where both 
kinetic terms vanish. It was shown in \cite{Seiberg-Witten}
by Seiberg and Witten that at such points, the
low-energy dynamics is controlled by tensionless strings and the
tensor multiplet that is sourced by them; it is possible that the
infinite sequences of models are actually redundant descriptions of
some tensionless string theories, although these models would still be
unstable with small fluctuations of the dilaton.
It is also possible that these infinite families can be realized as an
infinite sequence of unstable string compactifications
\footnote{Thanks to John McGreevy for discussions on this point}.

Note that we are interested here in theories which are coupled to
gravity with a finite Newton constant.  Thus, in string theory these
should be described by true string compactifications.  If we relax
this constraint, there is a much greater freedom in the construction
of arbitrary six-dimensional quantum field theories using ``local''
string constructions in terms of gauge fields on D-branes without
gravity. The infinite family of $SU(N)\times SU(N)$ theories constructed by
Schwarz can be realized by $N$ D5-branes at a $\Z_2$ singularity
\cite{intriligator, Blum2}. While this is a perfectly consistent family of gauge
theories, it cannot be coupled to gravity without pathologies.

The apparent inconsistency of the known infinite families of
anomaly-free theories in 6D supplies further evidence in support of
the hypothesis of string universality in six dimensions, which we
discuss in more detail in Section~\ref{sec:conjecture}.

\subsection{On the existence of infinite families}
\label{sec:existence-infinite}

In the previous subsection it was demonstrated that in the class of
theories with gauge group $SU(N)\times SO(M)$ with $A$ hypermultiplets
in $\mbox{Anti }SU(N)$ and $B$ hypermultiplets in $(N,M)$, there are
only a finite number of anomaly-free models which seem consistent.
Although some infinite families have been found which apparently
satisfy anomaly cancellation, these families exhibit pathologies which
seem to rule them out as consistent gravity theories.  It is
interesting to ask whether anomaly cancellation alone is a sufficient
condition to impose a finite constraint on the rank of the gauge group
appearing in any consistent supergravity theory in six dimensions.  It
would be hard to realize gauge groups of arbitrarily large rank in a
string compactification. A schematic argument for this is the fact
that gauge groups come from D-branes whose R-R charge must be
neutralized by negative tension objects like orientifold planes
.  Since these have a fixed negative
charge, the number of D-branes that can be packed into a compact space
is bounded from above, and so is the rank of the gauge group.  Note
that this argument is not airtight.  In four dimensions, for example,
D-branes in models with several R-R tadpoles can have negative charge
with respect to one or more of these tadpoles, making the proof of
finiteness of the number of available models a rather nontrivial
challenge \cite{Douglas-Taylor}.  It seems, however, in six
dimensions, that the range of possibilities is sufficiently small that
it is highly unlikely that any family of string compactifications
giving gauge groups of arbitrarily high rank is possible.

Let us consider as an example the gauge group $SU(N)$ with $x_i$
multiplets in the representation $R_i$ of the gauge group. We look for
the existence of an infinite family of theories with increasing rank,
without any restrictions on $x_i$ and $R_i$.  Anomaly-free models with
gauge group $SU(N)$ were also studied in \cite{Avramis-Kehagias},
though with only limited possible matter representations.

The absence of gravitational anomalies requires $n_H-n_V=244$. For the
theories in this class
\begin{equation}
\sum_i x_i \dim R_i - N^2 = 243
\end{equation}
If there exists an infinite family, it follows that 
\begin{equation}
\mathop{\rm lim}_{N\rightarrow \infty} \left( \sum_i x_i \dim R_i - N^2 \right)
\rightarrow {\mathcal O} (1)
\label{eq:limit}
\end{equation}
This condition restricts the possible representations $R_i$ to be
among
the 
fundamental, two-index symmetric, two-index anti-symmetric and adjoint
representations, since
any other representation has a dimension of order $\O(N^3)$. In fact, the only
possibilities for the matter hypermultiplets consistent with
(\ref{eq:limit})
are
\begin{enumerate}
\item 1 Adjoint + 244 neutral
\item 1 Symmetric + 1 Anti-symmetric + 243 neutral 
\item $N$ Fundamental + 243 neutral
\item 2 Anti-symmetric +  $N$ Fundamental + 243 neutral
\item 1 Symmetric +  $(N -1)/2$ Fundamental + 243 neutral, $N$  odd
\item 1 Anti-symmetric +  $(N +1)/2$ Fundamental + 243 neutral, $N$
  odd
\end{enumerate}
In cases 1 and 2, the anomaly polynomial is just $I=(\tr R^2)^2$ and
the kinetic term is zero. We discard these families as the gauge fields in these
theories have no dynamics. In the remaining cases, the $\tr F^4$ term
is present and nonvanishing for all $N$, and hence all such theories
are anomalous.

We have thus shown that there exist only finitely many anomaly-free
supersymmetric gauge theories with gauge group $SU(N)$ coupled to
gravity in six-dimensions.  A partial list of these theories with
restrictions on the matter content can be found in
\cite{Avramis-Kehagias}.  The absence of infinite families implies
that there is a maximal rank beyond which low-energy gauge theories
with gauge group $SU(N)$ cannot be consistently coupled to
supergravity in six dimensions.  Though we do not have explicit string
theoretic realizations of all theories in this class\footnote{In
particular, three and higher index representations of $SU(N)$ would be
hard to realize in a perturbative string compactification.}, the fact
that the gauge group rank is bounded from above, provides additional
evidence for string universality in six dimensions.  It would be
interesting to find a general proof of a bound on the rank of the
gauge group for arbitrary forms of the gauge group in six-dimensional
supergravity theories purely based on anomaly cancellation arguments.

\section{String Universality}
\label{sec:conjecture}

\begin{figure}
\begin{center}
\begin{picture}(200,100)(- 100,- 50)
\put(0,5){\makebox(0,0){\includegraphics[width=3in]{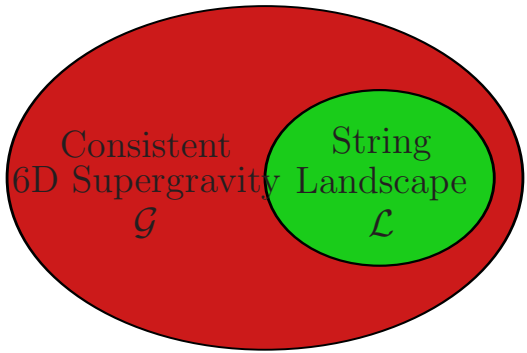}}}
\put(-45, 18){\makebox(0,0){\large Consistent}}
\put(-45, 3){\makebox(0,0){\large 6D Supergravity}}
\put(-45,-12){\makebox(0,0){\large ${\mathcal G}$}}
\put(44,18){\makebox(0,0){\large String}}
\put(44,3){\makebox(0,0){\large Landscape}}
\put(44,-12){\makebox(0,0){\large ${\mathcal L}$}}
\end{picture}
\end{center}
\caption[x]{\footnotesize The set of consistent
six-dimensional gravity theories
  ${\mathcal G}$ contains the set of models realized through
compactification of string theory to six dimensions ${\mathcal L}$.  The
conjecture of string universality
states that ${\mathcal G} ={\mathcal L}$.  If all models in ${\mathcal
  G}$ can be identified by features of the low-energy Lagrangian, then
the swampland can be identified as
${\mathcal S}={\mathcal G} \setminus{\mathcal L}$,
and in this situation string universality would imply that the
 6D SUSY swampland is empty (${\mathcal S} =\{ \}$).}
\label{f:gl}
\end{figure}

The close correspondence between the anomaly cancellation conditions
in six dimensions and the conditions governing the tadpole constraints
and matter field content of string compactifications suggests that the
set of consistent six-dimensional (1, 0) supersymmetric theories of matter
coupled to gravity may be identified almost completely through anomaly
cancellation conditions. Furthermore, this set of consistent models may be
essentially the same as the set of supersymmetric six-dimensional
models which can be realized through string compactifications.  In
this section we attempt to formulate a precise statement of how this
correspondence may be realized.  We denote the set of consistent
six-dimensional supergravity theories by ${\mathcal G}$, and the set of
models which can be realized through some string compactification
mechanism by ${\mathcal L}$.  These sets of theories are defined more
precisely in Subsections \ref{sec:g}, \ref{sec:l}.  Since all known
compactifications of string theory to six dimensions give apparently
consistent theories of quantum gravity, we assume ${\mathcal L}
\subseteq{\mathcal G}$.  The associated Venn diagram is depicted in
Figure~\ref{f:gl}.
The conjecture of string universality states that ${\mathcal G} ={\mathcal
  L}$.  The conjecture is stated more precisely for the limited class
of theories on which we focus in this paper in Subsection
\ref{sec:conjecture-statement}, where the related swampland picture of
\cite{swampland} is also discussed.  The current situation regarding
evidence for string universality in more general classes of theories,
including dimensions other than six, is discussed in
\ref{sec:extending}.  Connections to the question of the predictivity
of string theory are discussed briefly in \ref{sec:predictivity}.

\subsection{${\mathcal G}$:
Supersymmetric field theories with gravity in six dimensions}
\label{sec:g}

Supersymmetric gravity theories in six dimensions have a rich
structure and have been widely studied in the literature
\cite{6D-theories, 6D-anomalies}.  Much of the richness of this
structure arises in theories with multiple massless tensor multiplets 
\cite{Sagnotti-others, Sagnotti, o2}
and in limits with tensionless string excitations
\cite{Seiberg-Witten}.  For simplicity, in this paper we focus on
theories with a Lagrangian description, for which $n_T = 1$.  Thus, we
are interested in theories of six-dimensional gravity with one 6D
gravity SUSY multiplet, one tensor multiplet, and any number of
vector- and hyper- multiplets $n_V, n_H$ associated with an arbitrary
gauge group and matter content in any representations of the gauge
group.  While such theories are generally trivial in the infrared, we
can study many of their properties from the supersymmetric Lagrangian
description.  In particular, from the structure of the gauge group and
matter content we can analyze the anomaly structure of any given
theory.

We define ${\mathcal G}$ to be the set of consistent six-dimensional
supersymmetric quantum theories of gravity with gauge symmetry and
matter fields.  This definition requires some discussion.  We do not
{\it a priori} have a systematic way of identifying which 6D
Lagrangians are associated with consistent quantum theories of
gravity.  It is possible that string theory is the only way in which
gravity can be consistently quantized in any dimension; some
researchers in the field may take this as a working hypothesis, and in
some sense the notion of string universality has been implicit in much
work over the last several decades.  It is, however, logically
possible that string theory is a consistent quantization of some
theories of gravity, but that there are other UV-complete quantum
theories of gravity disjoint from the string theory landscape.  Thus,
suggesting that all consistent supergravity theories are described by
string theory is not tautological, but is a hypothesis subject to
mathematical proof or refutation.  To prove such a hypothesis, a key
step is finding criteria for determining which models lie in the set
${\mathcal G}$, independent of string theory.  This means that we wish to
find a set of conditions determining ${\mathcal G}$ which can be analyzed
from the point of view of the low-energy Lagrangian.  One of the main
points of this paper is that in the case of six-dimensional
supersymmetric theories we may be very close to having a complete set
of such conditions.  As discussed above, anomaly cancellation gives a
very strong set of conditions constraining the set of theories in
${\mathcal G}$.  Some models which satisfy local anomaly cancellation
however, may still not admit a UV completion to a quantum theory of
gravity.  For example, global anomalies must also be taken into
account \cite{Bershadsky-Vafa}.  For some models, extra massless
fields may be required in order to satisfy unitarity.  Or nontrivial
topological states may be required for consistency of the theory.  In
this case, an additional condition for consistency of the quantum
theory would involve the cancellation of anomalies on the world-volume
of the topological excitation, as suggested in \cite{Uranga}.  Other
constraints on the low-energy Lagrangian required for consistency of
supergravity theories in the UV are found in \cite{Allan-Nima}.  Thus,
while we do not claim to have a complete set of criteria by which
models in ${\mathcal G}$ can be determined, we suggest that at least for
six-dimensional supersymmetric theories, anomaly cancellation provides
a strong litmus test, and that further development of consistency
conditions based on low-energy considerations may lead to a complete
determination of what models may admit consistent quantization
independent of string theory considerations.

This discussion is closely related to the perspective presented by
Vafa in \cite{swampland}, developed further in \cite{swampland-2}.  In
that work, the ``swampland'' is defined to be the set of
consistent-looking effective field theories which are actually
inconsistent.  The implicit assumption in this work seems to be that
any theory not realized in string theory is inconsistent, so that the
swampland represents all theories which are consistent with what we
know about the low-energy theory, but which cannot be realized in
string theory.  The extent of the swampland thus depends upon the
power of our tools with which to probe the consistency of a theory
from its low-energy description.  In this context, the point of the
present work is to suggest that in the case of six-dimensional
supergravity theories, we may be close to having a complete
characterization of which low-energy theories are consistent.  Instead
of defining the space ${\mathcal G}$ of consistent gravity theories, we
could have chosen to define a space ${\mathcal C}$ of ``apparently
consistent'' theories of 6D gravity.  The swampland would then be
defined as the set ${\mathcal C} \setminus{\mathcal L}$.  The difficulty with
this definition is that it depends upon the set of tools we allow for
testing consistency of a theory.  If we restrict to anomaly
cancellation, it would define a particular class of theories ${\mathcal
C}$.  There are, however, other criteria, including those mentioned above, which may be used to refine
the set of apparently consistent theories based on the low-energy
Lagrangian.  The space ${\mathcal C}$ is therefore somewhat of a moving
target, which will be reduced in scope as our tools for discovering
quantum inconsistencies are improved, although we will always have
${\mathcal C} \supseteq{\mathcal G}$.  For purposes of clarity, therefore, we
have chosen to frame our discussion primarily in terms of the space
${\mathcal G}$, which represents a mathematically idealized static set,
despite our current ignorance of how to determine this set.  The goal
is then to find a set of criteria which are sufficiently stringent to
define a set of ``apparently consistent'' models ${\mathcal C}$ which
coincide with ${\mathcal G}$.  Our discussion here is therefore similar in
philosophy if not in semantics and notation, to the approach taken in
\cite{swampland, swampland-2}, where additional criteria based on
finiteness of moduli space \cite{Douglas-finite} and numbers of fields
are suggested as ways in which the set ${\mathcal C}$ can be reduced.

\subsection{${\mathcal L}$:
The six-dimensional SUSY string landscape}
\label{sec:l}

There are many ways to construct six-dimensional supersymmetric vacua
in string theory.  These include heterotic/type I compactifications on
K3 \cite{gsw}, F-theory compactifications \cite{Kachru-Vafa, Vafa-f,
  Morrison-Vafa, Sen, Aspinwall-Morrison}, and orbifold
constructions \cite{6d-orbifolds, o2}.  All these constructions are related through
various duality symmetries (reviewed in \cite{Aspinwall-review}).  It is quite possible that string
constructions exist which are related in some limits through duality
to constructions which have already been studied, but which include
new classes of vacua not yet considered.  By ${\mathcal L}$ we mean the
complete space of compactifications in the entire string theory
duality web.  Since string theory is not yet well-defined in a
background-independent fashion, we do not yet have a complete
mathematical definition of ${\mathcal L}$.  Nonetheless, if we assume that
there is some as-yet unknown definition of string theory we can assume
that there is some well-defined, yet unknown definition of the set
${\mathcal L}$.  Despite our ignorance about the general theory, we can at
any time define a set ${\mathcal K}$ of known string theory constructions.
Like the set ${\mathcal C}$ defined in the previous subsection, ${\mathcal K}$
is a moving target, which always satisfies ${\mathcal K} \subseteq{\mathcal
  L}$.  The point we would like to emphasize here is that in six
dimensions with supersymmetry, we may be close to understanding a wide
enough range of string constructions that realizing ${\mathcal K} ={\mathcal
  L}$ may be within reach.

\subsection{Statement of conjecture}
\label{sec:conjecture-statement}

The statement of  string universality in the class of theories of
interest in this paper can be stated as the following conjecture.

\vspace*{0.05in}
\noindent
{\bf Conjecture}: {\it All ${\mathcal N} = (1, 0)$ supersymmetric theories
  of gravity coupled to gauge fields and matter in six dimensions {\rm
  (}with one gravity and one tensor multiplet{\rm )} which are anomaly-free and
  consistent quantum theories can be realized through a compactification of
  string theory.}
\vspace*{0.05in}

In terms of the sets of theories defined in the previous two
subsections, this conjecture states that for a restricted class of
six-dimensional supergravity theories
\begin{equation}
{\mathcal G} ={\mathcal L} \,.
\label{eq:conjecture}
\end{equation}

The purpose in formulating this conjecture is to provide a concrete
framework in which progress can be made in understanding the
landscape of string vacua (${\mathcal L}$), consistent quantum gravity
theories (${\mathcal G}$), and the relationship between these classes of
theories.  As discussed above, both ${\mathcal L}$ and ${\mathcal G}$
represent some mathematically idealized sets which we do not yet know
how to completely determine.  We have tools for demonstrating that
certain theories are inconsistent, leaving a set ${\mathcal C}$ of
theories, as defined above, which are ``apparently consistent''.  We
have a set ${\mathcal K}$ of known string constructions.  The relationship
between these four sets is given by
\begin{equation}
{\mathcal K} \subseteq{\mathcal L} \subseteq{\mathcal G} \subseteq{\mathcal C} \,.
\end{equation}
Even though we do not know precisely how to define the sets ${\mathcal L}$
and ${\mathcal G}$, we have definitions of ${\mathcal K}$ and ${\mathcal C}$ which
over time form better bounds on the sets ${\mathcal L}$ and ${\mathcal G}$.
One way to prove the conjecture that ${\mathcal L} ={\mathcal G}$ is to prove
that ${\mathcal K} ={\mathcal C}$ since
\begin{equation}
{\mathcal K} ={\mathcal C} \; \; \;\Rightarrow
\;\;\;
{\mathcal L} ={\mathcal G}\,.
\label{eq:implication}
\end{equation}
The point we would like to make here is that in six-dimensional
supergravity theories, our definitions of ${\mathcal K}$ and ${\mathcal C}$
are sufficiently good that we are close to having ${\mathcal K} ={\mathcal
C}$, from which the stated conjecture would follow from
(\ref{eq:implication}).

As stated above, the purpose of making this conjecture is to provide a
framework for refining our understanding of the string landscape and
consistent gravity theories in an arena where short-term progress may
be possible.  In Section \ref{sec:uo} we considered a simple class of
6D SUSY models with gauge groups $SU(N) \times SO(M)$.  The analysis
carried out in that subsection proves that for this class of models
${\mathcal C} ={\mathcal K}$ and therefore the conjecture of string
universality holds in this limited class of models (with one
uncertain exception which we expect also has a string realization and
agrees with the conjecture). Here ${\mathcal C}$ can be defined as the set of
supersymmetric, anomaly-free 
models, with the stated limitations on gauge group and matter content
restricting to the subclass of theories of interest.
We also
considered in \ref{sec:infinite} several infinite families of models
satisfying anomaly factorization, and showed that these models have
wrong-sign gauge kinetic terms and therefore do not represent counterexamples to
the conjecture.

There are several ways in which one might imagine making progress in
proving the conjecture (\ref{eq:conjecture}).  The most
straightforward is to consider models which do not have known string
constructions, but which are not anomalous ({\it i.e.}, which lie in
${\mathcal C} \setminus{\mathcal K}$).  For any such model $m$ there are three
possibilities ---
\begin{enumerate}
\item The model admits a string construction, so $m \in{\mathcal K}$ with an improved definition of ${\mathcal K}$.
\item The model is inconsistent, so $m \not\in{\mathcal C}$ for an improved
${\mathcal C}$.
\item The model is a consistent theory of gravity
but cannot be realized in string theory, so $m \in{\mathcal G} \setminus{\mathcal
L}$, providing a counterexample to the conjecture of string
universality.  
\end{enumerate}
Thus, analyzing such models provides a systematic way
of expanding/refining the definitions of ${\mathcal K}, {\mathcal C}$.  At
present there are few models known of this type.  Categorizing and
analyzing these models may be helpful in increasing our understanding
of the situation and approaching a demonstration of the conjecture
stated above.

It is also possible that more abstract considerations may be useful in
proving the conjecture (\ref{eq:conjecture}), at least in restricted
classes of theories.  As shown in \ref{sec:uo}, in some cases the
anomaly cancellation conditions in six dimensions are precisely
equivalent to conditions on field theories derived from string
compactifications.  Proving this in a more general context, perhaps
using the mathematics of index theories, might provide a deeper
insight into the structure of the string landscape.

\subsection{Extending the conjecture}
\label{sec:extending}

We have stated the conjecture (\ref{eq:conjecture}) in the restricted
context of six-dimensional (1, 0) supersymmetric theories with one tensor multiplet because
it is in this arena that we have strong constraints on allowed
theories of quantum gravity, which come close to defining the same set
of models which can be realized using known string compactifications.
One could imagine generalizing this conjecture in several directions.

For example, one could drop the restriction that $n_T = 1$.
Unfortunately, our understanding of theories with more than one tensor
multiplet in six dimensions is rather limited, outside the insights
which have come from string theory (some of which are described in
\cite{Seiberg-Witten, o2}).  Thus, our understanding on the field theory
side is rather limited.  Nonetheless, anomaly cancellation conditions
can still be determined based on the massless field content and
representations of the theory \cite{Sagnotti-others, Sagnotti}, so
progress may be possible in this direction without the need for
tremendous new insights.  

We have focused here on chiral theories with (1, 0) supersymmetry
because they are constrained by anomaly cancellation.  One could
extend the conjecture to non-chiral theories with extended
supersymmetry.  Additional consistency conditions, as discussed above,
would be needed to constrain these theories.

The analogue of the conjecture made here in 10 dimensions is a much
simpler story, much of which is well-known.  Without the Green-Schwarz
mechanism, anomaly cancellation in 10 dimensions basically rules out
all chiral supergravity theories other than the type IIB theory
\cite{Alvarez-Gaume-Witten}.  The Green-Schwarz mechanism
\cite{Green-Schwarz} makes possible not only ${\mathcal N} = 1$
supergravity coupled to the SO(32) gauge theory, but also the $E_8
\times E_8$ theory (which led to the discovery of the corresponding
string theory), and two additional theories with gauge group
$U(1)^{496}$ and $E_8 \times U(1)^{248}$.  These latter theories have
not been shown either to exhibit inconsistencies as quantum theories
of gravity, or realized through string theory.  (Thus, in the language
of the previous subsection, these theories lie in the set ${\mathcal C}
\setminus{\mathcal K}$.) The proof of the conjecture ${\mathcal G} ={\mathcal L}$
in 10 dimensions would involve showing that these two theories either
admit string constructions or are inconsistent.  It seems most likely
that these models are inconsistent \footnote{Thanks to Allan Adams and
Joe Polchinski for discussions regarding these models}, as also
suggested in \cite{swampland, Fiol}, though this remains to be proven.
Discovering a mechanism for inconsistency of these theories may also
be a useful step in proving the conjecture in six dimensions, where
analogous anomaly-free models arise.  In 9, 8 and 7 dimensions, the space of 
possible string compactifications is also much better
understood \cite{triples} than in six dimensions.
Thus, it may be interesting to attempt to prove string universality in
these higher dimensions, which could give lessons relevant for more
complicated lower-dimensional situations.

Clearly, the case of physical interest is that of four-dimensional
theories.  One could conjecture that string universality holds in
four dimensions, but this hypothesis is much less compelling than in six
dimensions given our current state of knowledge, even if one restricts
attention to the class of supersymmetric theories.  For one thing, since
there are no purely gravitational anomalies in four dimensions, the
constraints on allowed field theories are much weaker.  For another
thing, our understanding of general string compactifications to four
dimensions is much more limited than for six dimensions.  Beyond the
enormous number of relatively well-understood compactifications
available through Calabi-Yau manifolds with branes and fluxes,
F-theory compactifications on elliptically fibered Calabi-Yau 4-folds,
and less well-understood M-theory compactifications on $G_2$
manifolds, there are also compactifications on non-K\"ahler manifolds,
and perhaps also non-geometric manifolds, with which we are only beginning
to come to grips.  Thus, at this time in four dimensions, the
definition of the set of known compactifications ${\mathcal K}$, while
vast, is much smaller than the set of apparently consistent 4D
theories coupled to gravity ${\mathcal C}$.  This is the framework in
which the discussion and analysis of the swampland in \cite{swampland,
swampland-2} is most relevant.  We need to make a vast improvement in
our understanding of both the space of string compactifications and
the set of criteria for showing that theories of gravity are quantum
inconsistent in four dimensions to even begin to approach a situation
where ${\mathcal C} ={\mathcal K}$.  It may be that aspects of the finiteness criteria
and moduli space structure
discussed in \cite{Douglas-finite,swampland,swampland-2} play a very
useful role in refining our understanding of these questions in four
dimensions.  It may be that some analogue of anomaly cancellation
persists in four dimensions which can dramatically cut down on the
number of apparently consistent gravity theories.  
It may be that string universality holds for four-dimensional theories
with supersymmetry, but that supersymmetry breaking mechanisms lead
to a constrained subset of non-supersymmetric low-energy theories in
4D.  And it may be that in four dimensions, even for supersymmetric
theories, ${\mathcal G} \neq{\mathcal L}$.
In any case,
further study of all these questions is clearly of great value, but we
are far from the point at which there is significant evidence for
string universality in four dimensions.

\subsection{Predictivity}
\label{sec:predictivity}

At the heart of the current controversy about the string landscape is
concern about the predictivity of the theory.  If some dynamical
principle would pick out a unique string vacuum, corresponding to our
observed standard model of particle physics with some specific
high-energy behavior and UV completion, then string theory would be
completely predictive and determining this vacuum would in principle
allow us to determine precisely what physics to expect above the TeV
scale.  If, on the other hand, eternal inflation populates the
landscape of theories, so that each string solution is realized in
some regions of a vast ``metaverse'' \cite{landscape-Susskind, flux-review}, then if
the solutions of string theory are sufficiently dense in the space of
allowed low-energy field theories, predictions for physics well below
the Planck or compactification scale would be difficult or impossible.
It is possible that the dynamics of string cosmology may define a
natural measure on the space of string solutions, which would  favor some solutions over others.
Currently, however, we lack a mathematically complete or
background-independent formulation of string theory.  It is likely
that significant progress in this direction will be needed to
understand the cosmological measure on the string landscape.  In this
brief discussion, we describe the situation for predictivity in the
absence of such a breakthrough.

If it is correct, or even close to correct, that string universality
holds in six dimensions, then in this case we seem in some sense
to be in the worst possible situation vis a vis low-energy
predictions.  If every possible consistent theory can be identified
from low-energy considerations, and all of these theories can be
realized in string theory, then string theory would seem to have no
predictive power for low-energy physics without some deep new insight
into the structure of the theory.  It has been suggested
\cite{statistics} that statistical methods may be useful in analyzing
the distribution of string vacua, so that even with a large number of
vacua densely populating the space of low-energy field theories, some
specific values for parameters in the low-energy theory may be overwhelmingly favored by the vacuum statistics.  On
the other hand, if the special class of examples considered in
\cite{Kumar-Taylor} is typical, the situation may be that each
possible consistent low-energy theory is realized in a single way in
string theory, which may be essentially unique up to duality
symmetries.

If we were living in six dimensions, then
this would seem like a very awkward situation for string
theory.  It should be emphasized, however, that there is no
reason {\it a priori} why a theory of quantum gravity relevant at the
Planck scale of $10^{19}$ GeV should make any prediction for physics
at the scale of 1 TeV, 16 orders of magnitude below the quantum
gravity scale.  String theory is valuable as a
framework for describing quantum gravity.  If in fact,
string theory can be used to provide a UV completion of
essentially any low-energy theory whose coupling to quantum gravity
does not violate some basic principle like unitary via anomalies, this
can be seen as a strength of the theory.  There is a certain symmetry
and elegance about the notion of a quantum gravity theory which
provides for the production of essentially all possible low-energy
behaviors in some regime of the theory or region of the metaverse.

If indeed, string theory can give rise to such a wide range of
low-energy behavior that predictions at the TeV scale cannot be made
precisely, it may bother some scientists that this makes the theory
difficult to test.  But, on the other hand, this does not make the
theory any less likely to be correct.  It just makes it more difficult
to verify.  In fact, one interesting feature of the six-dimensional
theories we have considered in this paper is that they all contain a
massless antisymmetric tensor field $B_{\mu \nu}$.  These theories
therefore contain a class of topological string-like defects which are
charged under the $B$ field.  It may well be that unitarity of these
theories requires that such topological string excitations be included
in the theory as quantum degrees of freedom.  Because these string
excitations are essentially equivalent to the fundamental string, the
proof of string universality in six dimensions, along with a
demonstration of the necessity for dynamical strings in these theories
might show that not only does string theory represent a
valid UV completion of each theory in this class, but that this UV
completion would essentially be unique, and that all quantum gravity
theories in this class would by necessity be string theories.  Such an
argument cannot work in four dimensions, where not all supersymmetric
theories have an antisymmetric 2-form field, but some more general
argument along these lines, including the possibility of M-theory
compactifications and other more exotic arrangements, may be
plausible.

In any case, these rather philosophical speculations apply in the case
of six dimensions, where there is some evidence that string theory
gives rise to at least a good fraction of the consistent quantum
gravity theories with supersymmetry.  Where we live, in four
dimensions with no manifest supersymmetry, our understanding of the
constraints on allowed field theories and string constructions is
still so weak that there is room for a large gap between the space of
string constructions and viable low-energy field theories, which would
give predictions for low-energy physics.  In attempting to understand
this problem better in six dimensions, where we are closer to a
complete story, we will inevitably develop better tools for
understanding the analogous situation in four dimensions.  Even if, in
the end, string theory does not provide a strong constraint on which
apparently consistent low-energy field theories with gravity and
supersymmetry admit a UV completion, probing these questions may give
us more powerful tools for understanding which low-energy theories
admit a consistent coupling to gravity based on completely low-energy
considerations; such tools may themselves provide new and interesting
insights or predictions for low-energy physics.  It is also possible
that attaining a reasonably complete description of the range of
allowed string vacua in four dimensions ({\it i.e.} approaching ${\mathcal
  K} \rightarrow{\mathcal L}$) may be a goal which is attained more
readily than developing tools for determining the inconsistency of
quantum gravity theories in four dimensions ({\it i.e.} approaching
${\mathcal C} \rightarrow{\mathcal G}$).  In this case, string theory can
provide a key to understanding how observable physics may be
consistently extended to higher energies.  In any case, the notion
that string theory provides a more or less complete catalogue of all
possible low-energy theories compatible with quantum gravity, even in
six dimensions, suggests
that any new and unexpected phenomena found in experiments at higher
energies may be realizable in the string theory context, so that,
as has been the case for many years, string theory can provide a
valuable tool for understanding the range of possible mechanisms for
low-energy phenomena, such as supersymmetry, supersymmetric breaking
mechanisms, unification, flavor symmetry, and perhaps any new physics
at the TeV scale which will be discovered in coming years.

\section{Conclusions}
\label{sec:conclusions}

In this paper we have shown that cancellation of gravitational, gauge,
and mixed anomalies gives a sufficient constraint on six-dimensional
supersymmetric theories of gravity with gauge and matter fields that
in some cases all models consistent with anomaly cancellation admit a
realization through string theory.  We have ruled out a number of
infinite families of models which satisfy anomaly factorization, so
that the gap is rather small between the set of known 6D models
satisfying anomaly cancellation and the set of models realized through
string compactification.  We have conjectured that all consistent 6D
supergravity theories with Lagrangian descriptions can be realized in
string theory, and that this set of models can be identified from
low-energy considerations.

In framing this conjecture of string universality in six dimensions,
we hope we have provided a framework for concrete progress in
understanding the string landscape and consistency conditions for
low-energy field theories to be coupled to gravity.  Progress in
proving or disproving string universality in
six-dimensional theories, either through specific examples or general
more abstract arguments will be of great value not only in six
dimensions, but in developing our understanding of these questions for
the more difficult but more important physical case of four
dimensions.

\vskip 0.2in 
\noindent
{\bf Acknowledgements} -- We would like to thank Allan Adams, Michael
Douglas, Eric Fitzgerald, David Guarrera, Ken Intriligator, Nabil Iqbal, John McGreevy, Greg Moore,
David Morrison, Joe Polchinski, Vladimir Rosenhaus, Cumrun Vafa, and Edward Witten for
discussions and correspondence related to this work.  WT would like to
thank the KITP for hospitality during the formative stages of this
work.  Thanks also to the organizers and participants of the
conference ``Perspectives on mathematics and physics,'' celebrating
the 85th birthday of I.\ M.\ Singer and his legendary contributions to
mathematics and physics, where some of this work was first presented,
and particular thanks from WT also to I.\ M.\ Singer, for providing
support and encouragement over many years as well as some of the
inspiration for this work.  This research was supported by the DOE
under contract \#DE-FC02-94ER40818. This research was also supported in part by the National Science Foundation under Grant No. PHY05-51164.



\noindent

\bibliographystyle{my-h-elsevier}

\end{document}